\documentclass[11pt,aps,onecolumn,amssymb,nofootinbib]{revtex4}
\usepackage{amsmath, amsthm, amscd, amssymb}
\usepackage{graphicx, braket}
\usepackage{bm}
\usepackage{bbm}
\usepackage{epstopdf}

\begin{document}

\title{\bf Galaxy formation catalyzed by gravastars and the JWST, revisited } \bigskip

\author{Stephen L. Adler}
\email{adler@ias.edu} \affiliation{Institute for Advanced Study,
Einstein Drive, Princeton, NJ 08540, USA.}

\begin{abstract}
We have proposed that galaxy formation is catalyzed by the collision of infalling and outstreaming particles from leaky, horizonless astrophysical black holes, most likely gravastars, and based on this gave a model for the disk galaxy scale length.  In this paper we modify our original scale length formula  by including an activation probability $P$ for a collision to lead to nucleation of star formation.  The revised  formula extrapolates from early universe JWST data to late time data to within a factor of five, and suggests that galaxy dimensions should systematically get smaller as the observed redshift z increases. We also show that particles recycling through gravastars can lead to a reduction in the temperature of the surrounding gas, through a ``heat pump'' refrigeration effect.  This can trigger galaxy formation through enhanced star formation in the vicinity of the gravastar.
\end{abstract}

\maketitle
\section{What lies inside the light sphere?}

The Event Horizon Telescope (EHT) (2020) has established, by viewing Sgr A* in our galaxy, and M87* in a neighboring galaxy,   that for an astrophysical black hole of mass $M$, the minimal stable orbit for photons, called the ``light sphere'', lies at the radius $3M$.  So the compact objects viewed by the EHT have the expected exterior geometry of a mathematical black hole.  But because of the Birkhoff  uniqueness theorem (Birkhoff, 1923) for  spherically symmetric solutions of the Einstein equations, the object interior to the nominal horizon at radius $2M$, whether a true black hole or an exotic compact object (Cardoso \& Pani, 2019) such as a gravastar (Mazur \& Mottola, 2001, 2004), (Adler, 2022a, 2024) and (Adler \& Doherty, 2023)   mimicking a black hole, will have to high accuracy the same exterior geometry.  Thus a key experimental question remains:  What type of object lies inside the observed light sphere?

Two approaches have been suggested to attempt to resolve this question.  The first, reviewed in (Cardoso \& Pani, 2019),  uses as a diagnostic the ringdown gravitational waves emitted in mergers of two holes.  The form of the waves emitted may have a different structure if the holes are true mathematical black holes with an event horizon, as opposed to the case in which the holes are horizonless compact objects.  A second approach (Adler, 2022b) notes that if astrophysical black holes have no event horizon or apparent horizon, as is the case with gravastars, then they will be ``leaky'', and interior particles will be able to exit.  This can influence astrophysical processes, such as young star formation near  Sgr A*  (Adler \& Singh, 2021), long time delays in black hole ejection of material from swallowed stars (Cendes, 2022), and galaxy formation  (Adler, 2022c), with the latter now discussed in more detail.

\section{Leaky astrophysical black holes as catalysts for galaxy formation, and elaboration of our original model to take cooling into account }

In a recent Gravitation Essay (Adler, 2022c), we have proposed that horizonless astrophysical black holes can act as the catalysts for galaxy formation.  This suggestion was originally motivated by a novel model for dark energy \big(reviewed in  (Adler, 2021a)\big)   that suppresses formation of black hole event horizons (Adler \& Ramazano\u glu, 2015) and apparent horizons (Adler, 2021b, 2022b) in the pure gravity, matter-free case, because of the factor $g_{00}^{-2}$ in the novel action, which would become infinite at a horizon where $g_{00}$ vanishes. Horizon suppression suggests that black holes may be leaky, which could have interesting astrophysical consequences, but the smallness of the cosmological constant which appears as an overall factor in the novel action poses a problem for robust astrophysical applications.  With this in mind, we more recently have studied (Adler, 2022a, 2024),
(Adler \& Doherty, 2023)  black hole models with nonzero matter content, focusing on a modification of
Mazur-Mottola gravastars (Mazur \& Mottola, 2001, 2004), which are an important exemplar of exotic compact objects. Our  version of what we call ``dynamical gravastars''  is based on using the Tolman-Oppenheimer-Volkoff (TOV) equation \big(for an exposition see (Zeldovich \& Novikov, 1971)\big), in which the  pressure is continuous, and the jump to a Gliner (Gliner, 1965) ground state, with pressure plus density summing to near zero, is through a jump in the energy density.   We found that even with zero cosmological constant, this model gives a gravastar with a metric  that approximates an exterior Schwarzschild solution to high accuracy, but with the metric component $g_{00}$  remaining always positive, and taking  small to very small values in the interior.  This gives a second mechanism for suppression of horizon formation, arising from properties of the TOV equation for relativistic matter,  quite distinct from the mechanism arising from the scale invariant dark energy action.

In a dynamical gravastar with the novel dark energy action  included, as modeled in (Adler, 2021a),  in the absence of accretion there is a very small black hole leakage or ``wind'' driven by the  cosmological constant, which would evaporate a $10^8\,M_\odot$ mass gravastar in $\sim 10^{23}$ years, corresponding to a mass loss rate so small that it will play no role in observed astrophysical processes such as galaxy formation. But even with zero cosmological constant, a much larger black hole wind can arise from accreting particles which exit the gravastar, on a time scale that depends primarily on the gravastar mass and the accreting particle impact parameter relative to the center of the gravastar (Adler, 2022a, 2024), (Adler and Doherty, 2023).   This gives a concrete model in which the black hole catalysis mechanism of (Adler, 2022c) can be realized, with negligible release of the gravastar binding energy.  The mechanism allows simultaneous  gravastar growth along with growth of the galaxy, since a fraction of accreting particles may not get back out of the gravastar, as a result of interactions with the gravastar interior medium, which at present cannot be calculated a priori.   This scenario is supported by the recent observation (Izumi et al., 2023) that in supermassive black hole feeding observed on subparsec scales, only $3\,\%$  of the accreted material is retained by the black hole, with the remaining $97\, \%$ fed back to the host galaxy.  If this observation is confirmed with others, it appears that the  fraction of accreting particles that get back out may be very substantial, as needed for our proposed galaxy formation mechanism.

A peculiar feature of galaxies is the existence of a diffuse structure that unfolds over many decades of length scales.  The model of (Adler, 2022c) ties this to the recent observation that nearly all galaxies have supermassive astrophysical black holes/gravastars at their center, by suggesting that supermassive  astrophysical black  holes form {\it before} galaxies in the early universe as a result of dust collapse. We thus incorporate a suggestion that supermassive black holes precede galaxies, made many years ago by Vestergaard (Vestergaard, 2004a,b). We propose that  these supermassive holes/gravastars
seed the formation of stars that constitute the galaxy, through collisions of   outstreaming with infalling particles.

To elaborate, instabilities in a gas cloud leading to localized collapse and star formation are governed by the Jeans criterion:  perturbations are unstable if the perturbation wave length exceeds the Jeans length (Binney \& Tremaine, 2008).  For our purposes here, it is useful to consider the  ``Jeans mass'', defined as the mass contained within a sphere of diameter the Jeans length.    Since the Jeans mass scales as $(T^3/n)^{1/2}$ (Weinberg, 2020),   with $T$ the gas temperature and $n$ the gas density, the Jeans mass will be decreased and star formation enhanced if either the gas density  $n$ is increased, the gas temperature $T$ is lowered, or both.  Collisions of  outstreaming with infalling particles around the gravastar can contribute to lowering the Jeans mass in two ways:
  First,  the collisions will lead to increased localized turbulence, with consequent overdensity regions, which by increasing the local density $n$ will reduce the Jeans mass and trigger instabilities that lead to star formation.  Second, and not discussed in our initial paper (Adler, 2022c), since the interior density of the gravastar/black hole is high \big(upwards of  $10^{21} {\rm nucleon}/{\rm cm}^3$  for supermassive holes (Chemlibre, 2024)\big), particles recycling through the gravastar interior will collide  with the internal matter in the gravastar, resulting in thermalization of motions transverse to the radial direction and decreased temperature for the exiting flow as it climbs out of the gravastar gravitational potential. (See the Appendix for further details.)  This will cool the surrounding gas in the exterior region within several collision lengths of the gravastar, and so by decreasing $T$ will reduce the Jeans mass and also contribute to instabilities that lead to star formation. This cooling of the gravastar exterior will be accompanied by a corresponding increase in the internal temperature of the gravastar. In other words, the gravastar is acting as a ``heat pump'', cooling the gas in its immediate surroundings by pumping heat to the gravastar interior, and thus creating favorable conditions for star and hence galaxy formation in the vicinity of the gravastar exterior.   Once the  gravastar heats up to point that the exiting flow is warmer than the surrounding gas, this cooling mechanism for the exterior gas will cease to operate, and star formation will be suppressed. If the gravastar can radiate heat back to its exterior at a sufficient rate to cool off again, there could be alternating cycles of star formation and its suppression.  \big(A recent observational study of the role of cooling in star formation is given in (MIT, 2024).  For an alternative proposal suggesting galaxy formation accelerated by dark matter primordial black holes, see Liu and Bromm (2024).\big)

If $\ell$ is the collision length of the outgoing particles (taken as neutral hydrogen), then the number (per unit solid angle per unit radius) of outstreaming particles at any radius from the central hole scales as $\exp(-r/\ell)$, corresponding to an approximately exponential scale length structure as observed for nearly all disk galaxies.  A similar estimate holds for any outwardly directed stream of particles, even if originating further out from the central hole, such as particles emerging from a galactic bulge.  The scale length estimate also applies whether the outstreaming particles are uniformly distributed over solid angle, or take the form of a number of distinct collimated jets.   A geometric estimate of the collision length is $\ell \simeq (A_H \rho_H)^{-1}$, where $A_H = \pi a_0^2$ is the cross sectional area of a hydrogen atom and $\rho_H$ is the density of atomic hydrogen in the region feeding galaxy formation, giving our original formula for the scale length,
\begin{equation}\label{original}
\ell \sim (\pi a_0^2 \rho_H)^{-1}~~~,
\end{equation}
where $a_0$ is the Bohr radius and $\rho_H$ is the density of atomic hydrogen in the proto-galactic region. Taking $\rho_H\sim 1 \,{\rm m}^{-3}$ for an initial estimate as done in (Adler, 2022c), this gives $\ell
\simeq 3.7\, {\rm kpc}$, close to  the observed value (Fathi et al., 2010) of $3.8 \pm 2.1 \, {\rm kpc}$, and suggests that the scale size of galaxies is fundamentally a property of atomic hydrogen.  Note that here, and throughout this paper,``size'' of a galaxy means linear dimension, not the mass of the galaxy.

Initially  the mechanism we are proposing will give rise to roughly spherical or more generally axially-symmetric galaxies, which then relax into disks through dissipation with conservation of angular momentum.    A purely geometric formula for the disk luminosity per unit area is obtained by projecting the three dimensional density $r^{-2}\exp(-r/\ell)$ onto a plane (Bosch, 2024). \big(In (Adler, 2022c) the spherical geometry prefactor $r^{-2}$ was omitted, which changes the power of disk radius in front of the projected density, but not the scale length in the exponential.\big) This gives a disk density profile $\Sigma(L) \propto L^{-1}\int_1^{\infty} dx x^{-1}(x^2-1)^{-1/2} \exp(-xL/\ell)$, which at large disk radius $L$ is proportional to $L^{-3/2}\exp(-L/\ell)$, and which at small $L$ diverges as $L^{-1}$, giving a convergent integral over the two dimensional disk. As noted to us by Tremaine (Tremaine, 2022), spherical density distributions and disk density distributions that are pure exponentials with the same scale length will have different total angular momenta, so angular momentum conservation dictates that they cannot have exactly the same form.  This is illustrated by the geometric projection model formula. In general the actual final disk profile resulting from relaxation of an initial sphere will depend on how angular momentum is acquired in the galaxy formation process.

Other  mechanisms have been suggested for producing an approximately exponential disk density profile, although without fixing the magnitude of the scale length $\ell$.  A mechanism based on maximizing entropy under angular momentum mixing by radial migration has been suggested by Herpich, Tremaine \& Rix (2017), and they give extensive references to earlier proposals. Since angular momentum has length dimension zero, an argument based solely on the angular momentum profile cannot by itself predict the dimension one disk scale length.  An additional assumption, introducing a parameter of dimension one that is fitted to the observed scale length $\ell$, is needed.  The mechanism of Herpich et al. (2017) suggests that an initially formed approximately exponential density profile, as in our proposal, would  be stable under subsequent galactic dynamics, and so can be complementary to our model.  In their analysis,  radial migration of stars in the disk plays an important role.  This can give a smearing of the exponential predicted by our model, but we believe it is very unlikely that this will change the scale length by more than a factor of order unity if at all; note in this regard that the projection  from three dimensional exponential scaling to  disk exponential scaling only changes the power of length in front of the exponential, but not the scale length in the exponent. Similarly, if the outgoing particles have velocities corresponding to positive energy at spatial infinity, the center of mass of a collision will not be at rest, again leading to a smearing of the exponential distribution.

\section{Problems with the original scale length formula, and a proposed extension}

Further examination, however,  indicates a number of problems with the suggested scale length formula of Eq. \eqref{original}.  First of all, the intergalactic medium (IGM) density relevant for the formula is not the average for the universe, which includes cosmic voids, but rather the  value for the Local Group in which our galaxy resides.  For the IGM density in the Local Group, Kahn and Woltjer (1959) \big(following on a suggestion of Spitzer (1956)\big)  have calculated that $\rho_H \sim 100 {\rm m}^{-3}$  is needed for dynamical stability \big(see also the more recent article (Rasmussen et al., 2003)\big).  However, Tremaine (2024) has pointed out to us that since Kahn and Woltjer (1959) were unaware of the large dark matter content of the universe, their estimate should be taken as giving the total matter density $\rho_m$, whereas the hydrogen density $\rho_H$ is roughly a factor of six smaller, that is $\rho_H \sim  17 {\rm m}^{-3}$.    Since this is a factor of 17 larger than the value used in our original formula, the corresponding scale length predicted by Eq. \eqref{original} becomes a factor of 17 smaller than observed.  And extrapolating from Eq. \eqref{original} with the revised  $\rho_H$ density estimate to  redshifts of $z=11$ and $z=13$, where the scale length should be reduced by  factors of $12^3= 1728$ and $14^3=2,744$ respectively, one gets predicted scale lengths of $\sim 0.13{\rm pc}$ and   $\sim 0.08{\rm pc}$. These are over three and a half orders of magnitude smaller than the size of two galaxies reported from JWST measurements by Naidu et al. (2022) at $z=11$ and $z=13$,  with scale lengths estimated as 0.7 kpc and 0.5 kpc respectively.

As a way or remedying these problems, we propose modifying Eq. \eqref{original} to
\begin{equation}\label{original1}
\ell \sim (P \pi a_0^2 \rho_H)^{-1}~~~,
\end{equation}
where $P$ is an activation probability for a collision of an outgoing particle with an ingoing one to nucleate star formation, through contributing to localized turbulence and overdense regions, and also to gas cooling, which enhance star formation via Jeans instabilities.    To estimate a posteriori what $P$ should be, we assume that in the early universe one can approximate the density of hydrogen atoms or ions as (WMAP, 2024)
\begin{equation}\label{hdens}
\rho_H \simeq 0.27 (z+1)^3 {\rm m}^{-3}~~~.
\end{equation}
Comparing with the z=11 galaxy observed by  Naidu et al.(2022) one gets a probability value
\begin{equation}\label{prob}
P\simeq 0.011~~~,
\end{equation}
more plausible than our original assumption of $P=1$.  Comparing Eq. \eqref{original1} with the late time observed disk galaxy scale length of $3.8 \pm 2.1 \, {\rm kpc}$, one gets a density estimate of $\rho_H \simeq (80 \pm 40) {\rm m}^{-3}$, a factor of five  larger than the Local Group IGM hydrogen density estimate of  $17 {\rm m}^{-3}$.  So the revised formula Eq. \eqref{original1} gives reasonable order of magnitude scale length estimates both at late times and in the early universe as observed by the JWST.  Note that we have focused on sale size regularities since these can have a purely kinematic, geometrical basis, whereas making predictions about the elapsed time for star and galaxy formation will require a detailed knowledge of rates of various contributing processes, which are not predicted by our model. Our simple model has focused on ballistic collisions with constant probability $P$.  Possible extensions include making $P$ a function of the particle streaming velocity, and/or including in the model diffusion effects resulting from particle collisions.

\section{Discussion}

A clear qualitative prediction of Eq. \eqref{original1} is that galaxy sizes should decrease systematically with increasing redshift, a trend evident in the JWST data indicating that galaxies in the early universe are smaller in linear dimension (but not in mass) than expectations based on the late time universe (Naidu et al., 2022), (Labb\'e et al., 2022), (Harikane et al., 2022), (Maiolino et al., 2024).  Our model also posits that copious supermassive black holes should be present in the early universe, again a trend appearing in early JWST data (Wood, 2023) , (Natarajan et al., 2023).  If black holes come first, and then galaxies are nucleated by them, another qualitative prediction is that the ratio of galaxy mass to central black hole mass should increase with cosmic time, which agrees  with recent JWST data (Eilers et al., 2024) \big(See also (Ferrarese (2022), Bandara, Crampton \& Simard (2009), and Trakhtembrot (2019).\big)  However, turning this into a quantitative prediction will depend on rates that we have emphasized are not predicted in our model.

Formation of supermassive black holes (SMBH) before galaxies requires a mechanism for SMBH creation in the early universe.  This will involve a collapse of initial visible matter, radiation, or dark matter in the high density, high temperature conditions of the early universe. Since  this involves a competition between the contrarily evolving temperature and inverse mass-energy density (recall that the Jeans mass scales as $(T^3/n)^{1/2}$ , detailed calculations are needed, and  self-interactions of the collapsing particles  may play an important role.  See for example (Balberg \& Shapiro, 2002) and (Feng et al., 2021)  for detailed models involving collapsing dark matter, and (Maiolino et al., 2024) for a discussion of the question of small versus large seeds for the collapse. To the best of our knowledge there is no definitive mechanism for SMBH formation in the very early universe.  Since in the dynamical gravastar calculations of (Adler, 2022a, 2024) and (Adler \& Doherty, 2023)   relativistic matter plays a preferred role, we believe it is important to investigate SMBH formation scenarios involving an initially relativistic gas of particles.

As noted, a salient feature emerging from JWST observations is that galaxies with high redshifts are smaller in diameter but brighter than anticipated, indicating higher stellar mass densities than had been expected on the basis of prior galaxy formation models. This has provoked discussion of whether the $\Lambda CDM$ standard model is valid at high redshifts (Boylan-Kolchin, 2022), (Lovell et al., 2022), or whether new cosmological ingredients are needed for early universe galaxy formation, such as novel forms of dark energy (Menci et al., 2022) or dark matter (Gong et al., 2022). The  mechanism for galaxy formation discussed in the preceding sections, utilizing collisions between infalling particles and outstreaming particles emerging from a central gravastar after accretion of matter, may give a way of providing a mechanism for  efficient stellar formation within the framework of the standard cosmological model, once a central SMBH has formed.    This mechanism can be tested by incorporating  an outgoing particle stream from a central source into  computer programs simulating galaxy formation in $\Lambda CDM$ cosmology, with parameterized energy, impact parameter, and flux distributions.

Finally, our model provides a mechanism for direct feedback from the central black hole to the growing host galaxy, which could play a role in understanding the $M$-$\sigma$ relation between the mass $M$ of the central hole and the velocity dispersion $\sigma$ of the stars in the central galactic bulge, complementing other mechanisms which have been proposed \big(see  (Ostriker, 2000), (King, 2003)\big).  Collisions of outgoing  particles from the central black hole (as well as infalling ones) with stars in the galactic bulge will result in heating, which will be reflected in velocity dispersion, and plausibly the larger the central hole the larger the heating and consequent dispersion.  However, a quantitative account will require rate estimates, and this is another important topic for further investigation.

{\it Added Note:} After this paper was posted to arXiv, I learned of a paper of Silk et. al. (2024) proposing a similar model for galaxy formation and black hole formation, also stimulated by JWST observations.

\section{Acknowledgement}
I wish to thank Fethi Ramazano\u glu for bringing the review (Cardoso \& Pani, 2019) to my attention, and to thank Scott Tremaine for helpful comments on drafts of both this paper and its predecessor.  I also wish to thank the referee for helpful comments and references.

\appendix
\section{Temperature Redshift}

 Tolman and Ehrenfest (1930) have shown that in a static gravitational field, the proper temperature $T$ of a system in equilibrium obeys the law $T g_{00}^{1/2}={\rm constant} $.  Since $kT$ is an energy, with $k$ Boltzmann's constant, this is an analog of the usual gravitational redshift.  For a gravastar, the calculations of (Adler, 2022a, 2024) and (Adler \& Doherty, 2023)  show that as one moves in from infinity, $g_{00}$ decreases from unity to an exponentially small value near the nominal horizon, with an even further exponential decrease as one moves into the gravastar interior.   Hence by the Tolman-Ehrenfest law, the temperature of a particle incoming from infinity is blue-shifted dramatically as it falls into the gravastar.  In general the internal temperature of the gravastar will be lower than the temperature attained by an infalling particle, since as a gravastar forms by dust collapse, the intermediate masses until its final size is reached will be characterized by less deep potential wells in $g_{00}$ than the well characterizing the final gravastar configuration.   So through collisions with the matter deep inside the gravastar, the particle will lose energy and lower its temperature.  As the particle moves back off to infinity, its  temperature is red-shifted by the same factor as it was blue-shifted falling in, but since it lost energy to the matter inside the gravastar, it will emerge at infinity with a lower temperature than it had when it started falling in.  This is how the gravastar can act as a ``heat pump'' or ``cosmic refrigerator'', lowering the temperature of the surrounding gas.  Since transverse motions that are damped out this way are equivalent to high angular momenta relative to the center of the gravastar, by acting in this way, gravastars can evade the argument given by Colgate and Petschek (1986) that ``There seems to be too much angular momentum in the universe to allow the formation of stars...~.''

\bigskip\bigskip
{\bf Additional information:}
The author has no competing interests, and there is no separate data to be made available.

\bigskip\bigskip
{\bf Author contributions:} This paper is entirely the work of  Stephen L. Adler.
\end{document}